\documentclass[prl,reprint,superscriptaddress,showpacs]{revtex4-1}
\usepackage{amsmath}
\usepackage{amssymb}
\usepackage{txfonts}
\usepackage{color}
\usepackage[unicode,breaklinks]{hyperref}
\hypersetup{
    unicode=true,
    a4paper=true,
    plainpages=false,
    pdftitle={Stochastic Growth Dynamics, Composite Defects and Metastability in Quenched Immiscible Binary Condensates},
    pdfauthor={I.-K. Liu, R. W. Pattinson, T. P. Billam, S. A. Gardiner, S. L. Cornish, T.-M. Huang, W.-W. Lin, S.-C. Gou, N. G. Parker, and N. P. Proukakis},
    pdfsubject={Stochastic Growth Dynamics, Composite Defects and Metastability in Quenched Immiscible Binary Condensates
},
    colorlinks=true,
    linkcolor=blue,
    citecolor=blue,
    filecolor=black,
    urlcolor=blue
}
\usepackage{amssymb}
\usepackage{graphicx}
\usepackage{upgreek}

\newcommand{\total}{d}

\begin{document}

\title{Stochastic Growth Dynamics and Composite Defects in Quenched Immiscible Binary Condensates
}

\author{I.-K. Liu}
\affiliation{Department of Physics, National Changhua University of Education, Changhua 50058, Taiwan}
\affiliation{Joint Quantum Centre (JQC) Durham--Newcastle, School of Mathematics and Statistics, \\ Newcastle University, Newcastle upon Tyne, NE1 7RU, United Kingdom}
\author{R. W. Pattinson}
\affiliation{Joint Quantum Centre (JQC) Durham--Newcastle, School of Mathematics and Statistics, \\ Newcastle University, Newcastle upon Tyne, NE1 7RU, United Kingdom}
\author{T. P. Billam}
\affiliation{Joint Quantum Centre (JQC) Durham--Newcastle, Department of Physics, Durham University, \\ Durham, DH1 3LE, United Kingdom}
\author{S. A. Gardiner}
\affiliation{Joint Quantum Centre (JQC) Durham--Newcastle, Department of Physics, Durham University, \\ Durham, DH1 3LE, United Kingdom}
\author{S. L. Cornish}
\affiliation{Joint Quantum Centre (JQC) Durham--Newcastle, Department of Physics, Durham University, \\ Durham, DH1 3LE, United Kingdom}
\author{T.-M. Huang}
\affiliation{Department of Mathematics, National Taiwan Normal University, Taipei 11677, Taiwan}
\author{W.-W. Lin}
\affiliation{Department of Applied Mathematics and Shing-Tung Yau Center, National Chiao Tung University, Hsinchu 30010, Taiwan}
\author{S.-C. Gou}
\affiliation{Department of Physics, National Changhua University of Education, Changhua 50058, Taiwan}
\author{N. G. Parker}
\affiliation{Joint Quantum Centre (JQC) Durham--Newcastle, School of Mathematics and Statistics, \\ Newcastle University, Newcastle upon Tyne, NE1 7RU, United Kingdom}
\author{N. P. Proukakis}
\affiliation{Joint Quantum Centre (JQC) Durham--Newcastle, School of Mathematics and Statistics, \\ Newcastle University, Newcastle upon Tyne, NE1 7RU, United Kingdom}
\email{nikolaos.proukakis@ncl.ac.uk}



\begin{abstract}\noindent
We study the sensitivity of coupled condensate formation dynamics  on the
history of initial stochastic domain formation in the context of
instantaneously quenched elongated harmonically-trapped immiscible
two-component atomic Bose gases. The spontaneous generation of defects in the
fastest condensing component, and subsequent coarse-graining dynamics, can lead
to a deep oscillating microtrap into which the other component condenses,
thereby establishing a long-lived composite defect in the form of a dark-bright
solitary wave.  We numerically map out diverse key aspects of these competing
growth dynamics, focussing on the role of shot-to-shot fluctuations and global
parameter changes (initial state choices, quench parameters and condensate
growth rates).  We conclude that phase-separated structures observable on
experimental timescales are likely to be metastable states whose form is
influenced by the stability and dynamics of the spontaneously-emerging
dark-bright solitary wave. 
\end{abstract}
\maketitle

{\em Introduction: }
Pattern formation and the presence of coexisting phases in spatially separated
domains are an emergent feature of diverse dynamical systems throughout physics
\cite{Cross1993a}, chemistry \cite{Meron1992a}, and biology
\cite{MurrayMathematicalBiology}.  Ultracold gases offer a highly controllable
theoretical and experimental test-bed for studying these phenomena,
particularly in the context of condensate formation dynamics
\cite{Stoof1997a,Kagan1997a,Semikoz1997a}, a subject of sustained interest and
significance in non-equilibrium physics, both in the ultracold gas context
\cite{Levich1978a, miesner_stamper-kurn_98, Gardiner1998a, bijlsma_zaremba_00,
stamper-kurn_miesner_98, stoof_bijlsma_01, kohl_davis_02,
proukakis_schmiedmayer_06, hugbart_retter_07, esslinger_07a, donner_ritter_07,
Garrett2011a, hadzibabic_15, Ronen2008a}, and beyond \cite{delValle2009a,
Nardin2009a, deLeeuw2013a, Chiocchetta}. Previous works have highlighted the
generic importance of the Kibble-Zurek (KZ) mechanism in defect formation
during phase transitions \cite{kibble_76,zurek_85}, connecting ultracold gases
with a range of phenomena from high-energy and condensed matter physics
\cite{Zurek1996177,Kibble2013a,delcampo_kibble_13}, with composite defects
\cite{kibble_njp} emerging in the context of multi-component fields by
condensation in the defect core \cite{antunes_gandra_06}.

\begin{figure}[b!]
\centering
\includegraphics[width=\columnwidth]{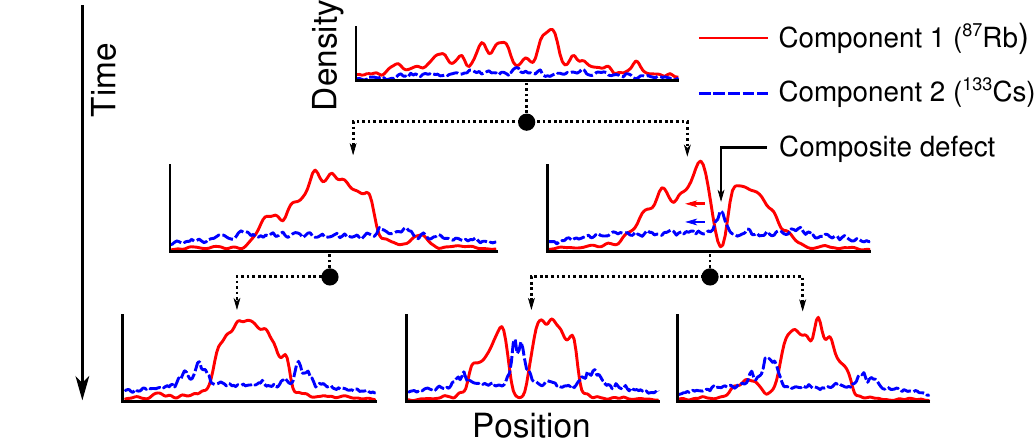}
\caption{(color online). Memory of stochastic defect dynamics in an immiscible
two-component condensate (schematic, compiled from simulation data). On short
post-quench time scales, the fastest-condensing component (here $^{87}$Rb)
contains multiple spontaneously-generated defects (top). The ensuing stochastic
dynamics generically lead to either destruction of defects, or survival of a
single composite defect in which the second component (here $^{133}$Cs)
preferentially condenses, forming a dark-bright solitary wave (middle row). The
long-lived metastable states (bottom
row) retain a memory of the prior stochastic dynamics. \label{fig:1}}
\end{figure}

Experiments with single-component atomic Bose-Einstein condensates (BECs) have
revealed the spontaneous formation of defects in the form of vortices
\cite{weiler_neely_08}, dark solitonic vortices
\cite{lamporesi_donadello_13,trento_prl} and persistent currents
\cite{corman_chomaz_14} during the BEC phase transition, providing quantitative
confirmation of KZ scaling
\cite{zurek_09,damski_zurek_10,witkowska_deuar_11,delCampo2011a,hadzibabic_15}.
In multi-component BECs (e.g.\ $^{87}$Rb-$^{41}$K \cite{Modugno2002a},
$^{87}$Rb-$^{85}$Rb \cite{Papp2008a} and $^{87}$Rb-$^{133}$Cs
\cite{McCarron2011a,lercher_takehoshi_11}), studies have focused on the
formation of two same-species BECs in a tunnel-coupled geometry \cite{Su2013a},
and quantum phase transitions in binary
\cite{Sabbatini2011a,Sabbatini2012a,Hofmann2014a} and spinor Bose gases
\cite{Sadler2006a,delCampo2011a,Saito2013a, Swislocki2013a, Witkowska2013a,
Stamper-Kurn2013a, De2014a}.  In many scenarios, however, potentially
non-universal post-quench defect dynamics strongly influence system behavior.
Work on homogeneous spinor Bose gases \cite{Swislocki2013a,Witkowska2013a} has
identified two distinct timescales in the process of domain formation: a short
timescale for initial domain formation via the KZ mechanism, and a long
``memory'' timescale for nonlinear coarse-graining dynamics to erase traces of
the initial domains.  Related work indicates that characteristic memory
timescales may exceed experimental system lifetimes \cite{Stamper-Kurn2013a,
De2014a}.

\begin{figure*}[t!]
\includegraphics[width=\textwidth]{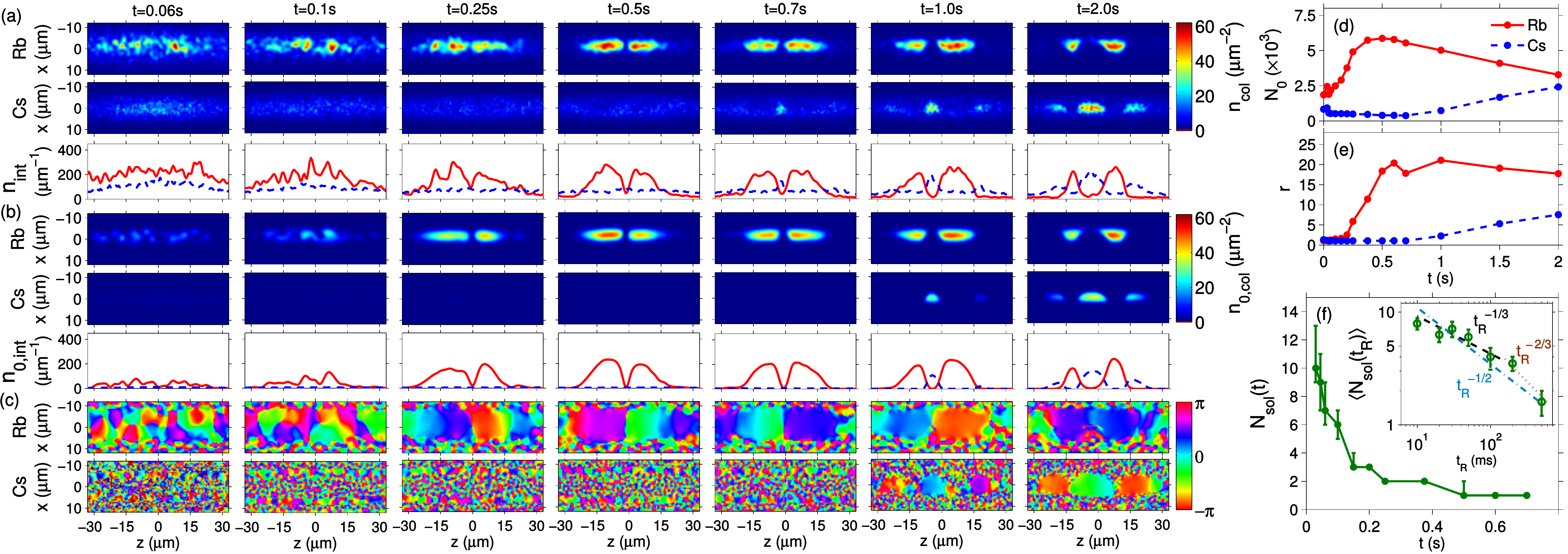}
\caption{(color online)
Typical numerical evolution of a quenched two-component system.  (a)--(b):
post-quench evolution, up to $2\,$s, of 2D column densities ($n_{\rm col}$,
$n_{\rm 0,col}$) and 1D integrated density profiles ($n_{\rm int}$, $n_{\rm
0,int}$) for c-field and PO condensate respectively. (c): corresponding
condensate phase. (d)--(e): evolution of (respectively) condensate number,
$N_0$, and occupation ratio, $r$, between condensate mode and the
next-most-highly occupied mode. (f, main plot): single-run evolution of the
number of spontaneously-generated dark solitons, $N_{\rm sol}$(t) (obtained, to
within the estimated uncertainty shown, by examing density minima and phase
jumps, similar to the procedure of \cite{damski_zurek_10}).  In all simulations
in this work, both components are confined in harmonic traps with frequencies
$(\omega_z^{\rm Rb}, \omega_{\perp}^{\rm Rb}, \omega_z^{\rm Cs},
\omega_{\perp}^{\rm Cs}) = 2\pi\times(3.89,32.2,4.55,40.2)$~Hz, with
longitudinal and transverse shifts of $\sim$1~$\upmu$m in their centers
\cite{McCarron2011a, Pattinson2013a}, and we use scattering lengths $(a_{\rm
Rb,Rb},a_{\rm Cs,Cs},a_{\rm Rb,Cs}) = (100,280,650)$ Bohr radii.  Temperature
is quenched from $T_0=80\,$nK to $T=20\,$nK.  In the first part of this work,
we use initial chemical potentials $\mu_{\rm Rb}/k_{\rm B}=2.13$~nK, $\mu_{\rm
Cs}=0.956 \mu_{\rm Rb}$ ($\mu_{\rm Rb}=1.38 \hbar \omega_{\perp}^{\rm Rb}$,
$\mu_{\rm Cs}=1.05 \hbar \omega_{\perp}^{\rm Cs}$), and rates $\gamma_{\rm
Rb}=\gamma_{\rm Cs}=0.263\,$s$^{-1}$ (equivalent to $\hbar \gamma_k/k_{\rm B}T
= 10^{-4}$), with $\mu_{\rm Rb}'=\mu_{\rm Rb}^{{\protect \phantom{\prime}}}$,
$\mu_{\rm Cs}'= 7.34 \mu_{\rm Cs}^{{\protect \phantom{\prime}}}$. The initial
condition is an equilibrated state with no single condensate mode ($r \sim
1$). In (f, inset) we also plot the average defect number $\langle N_{\rm sol}
\rangle$ immediately following \textit{finite-duration} quenches with ($T_k$,
$\mu_k$) ramped linearly from initial to final values (as given above) over time
$t_R$. Lines qualitatively show potential defect number scalings with $t_R$, 
rather than KZ predictions for our system (see text).
\label{fig:2}}
\end{figure*}

In this Rapid Communication we discuss the competing growth dynamics of
instantaneously-quenched immiscible two-component BECs, providing strong
evidence that the density profiles emerging over experimentally relevant
timescales are determined by the history of spontaneous defect formation, and
subsequent dynamics [Fig.~\ref{fig:1}]. This long-term memory is facilitated by
the added stability provided by formation, during coarse-graining dynamics,
of a composite dark-bright solitary wave \cite{Busch2001a, Ohberg2001a,
Shrestha2009a, Becker2008a, Hamner2011a, Hamner2013a} defect, a process found
to be robust to perturbations, but sensitive to shot-to-shot fluctuations and
global parameter changes.

While the equilibrium states of immiscible two-component BECs in the mean-field
approximation are known to display various symmetric and asymmetric structures
\cite{Trippenbach2000a, Pattinson2013a} similar to those observed in experiment
\cite{Papp2008a, McCarron2011a}, we show similar structures
emerging as part of a rich non-equilibrium dynamical behavior long
\textit{before} the system reaches an equilibrium state. Our work offers
insight into the complexities of two-component BEC formation, and suggests
caution when concluding that an experimental immiscible two-component BEC has
reached a true equilibrium state.

\begin{figure}[t!]
\includegraphics[width=\columnwidth]{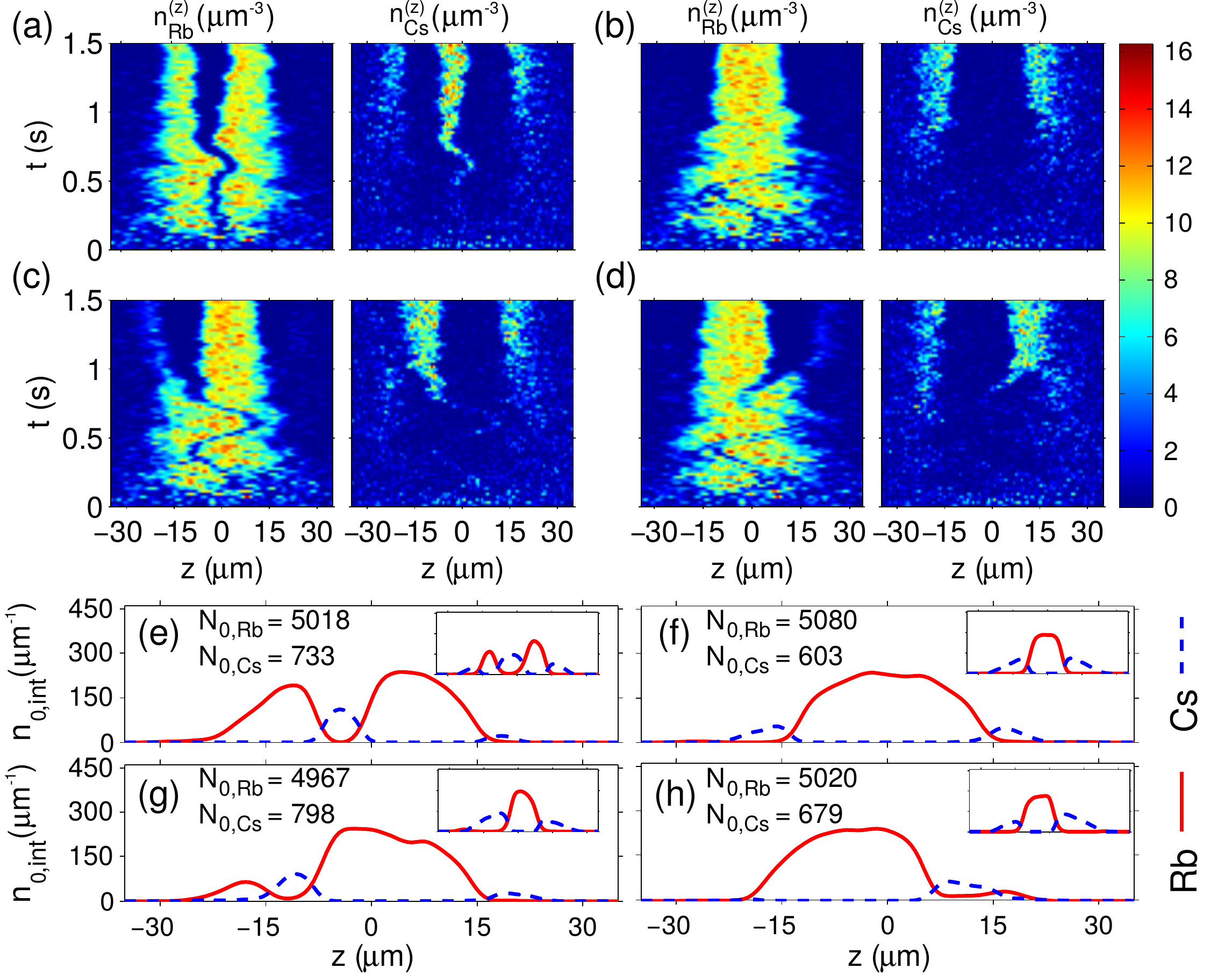}
\caption{(color online). Evolution for four different post-quench random noise sequences, with
identical pre-quench state. (a--d) show the c-field density evolution
[shown here via $n_k^{(z)}=n_k(x=0,y=0,z)$], and (e--h) the condensate density profiles after
$1\,$s (insets $2\,$s). After this time the Rb defect has either: been stabilized by Cs
infilling (e); fully decayed within Rb (f); or decayed to the left (g) or right
(h) edge of the condensate. Quench parameters are as in Fig.~\ref{fig:2}.
\label{fig:3}}
\end{figure}

\begin{figure}[t]
\includegraphics[width=\columnwidth]{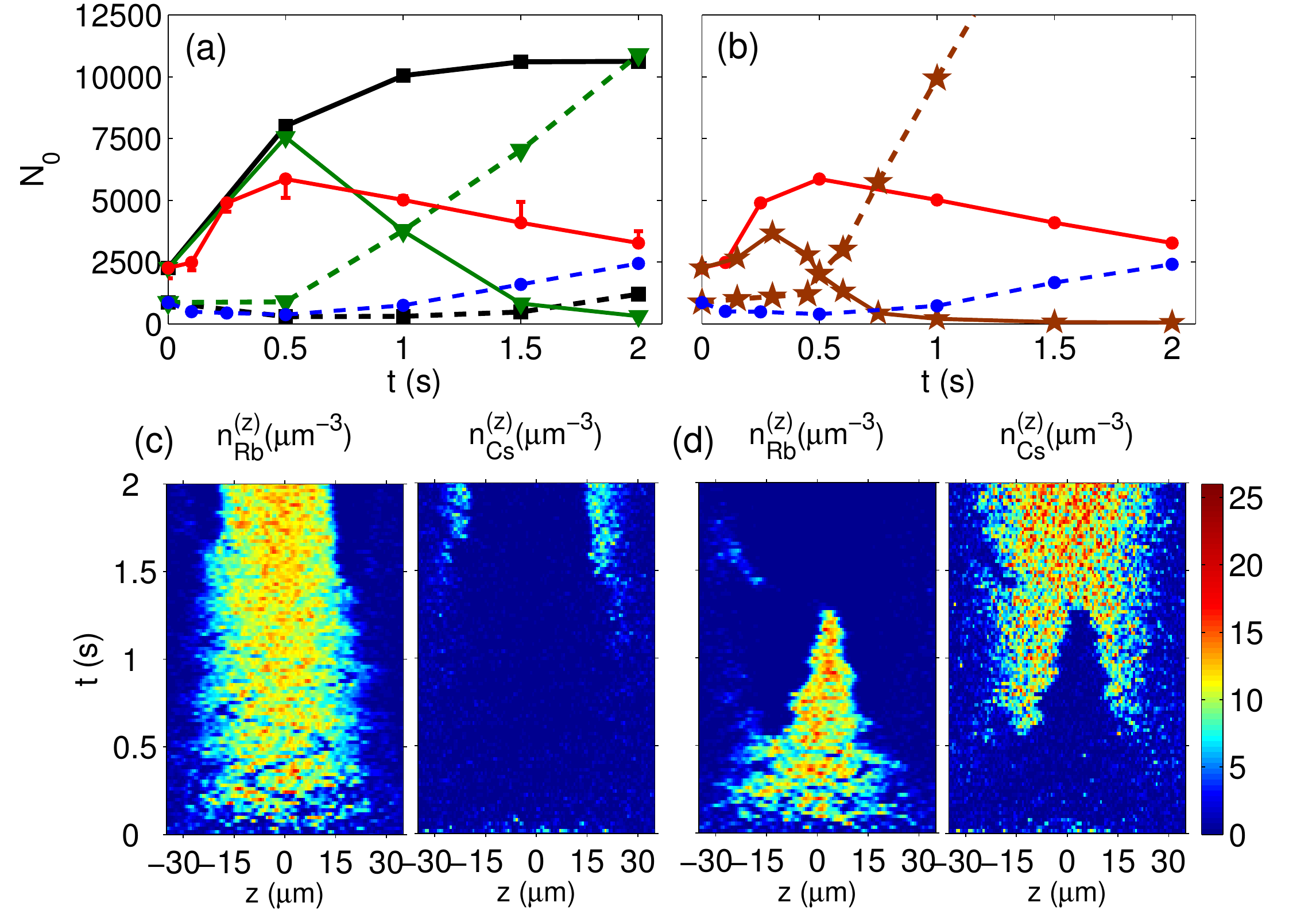}
\caption{(color online). Dependence of evolution on quench parameters. (a)
shows condensate number evolution (Rb solid, Cs dashed) for different final
chemical potentials $\mu_k'$; standard parameters $\mu_{\rm Rb}'= \mu_{\rm
Rb}^{\protect \phantom{\prime}}$, $\mu_{\rm Cs}'= 7.34 \mu_{\rm Cs}^{\protect
\phantom{\prime}}$ (red/blue circles); $\mu_{\rm Rb}'= 2.63\mu_{\rm
Rb}^{\protect \phantom{\prime}}$, $\mu_{\rm Cs}'= 7.34 \mu_{\rm Cs}^{\protect
\phantom{\prime}}$ [black squares, example c-field density evolution shown in
(c)] ; $\mu_{\rm Rb}'= \mu_{\rm Rb}^{\protect \phantom{\prime}}$, $\mu_{\rm
Cs}'= 11.9 \mu_{\rm Cs}^{\protect \phantom{\prime}}$ [green triangles, example
c-field density evolution shown in (d)]. Error bars in (a) for standard
parameters indicate the standard deviation over 6 numerical trajectories. (b)
shows condensate number evolution for different rates $\gamma_k$; standard
parameters $\gamma_{\rm Cs}=\gamma_{\rm Rb}=0.263\,$s$^{-1}$ (red/blue
circles), and  $\gamma_{\rm Cs}=10\gamma_{\rm Rb}=2.63\,$s$^{-1}$ (brown
stars).
\label{fig:4}}
\end{figure}

\begin{figure}[t]
\includegraphics[width=\columnwidth]{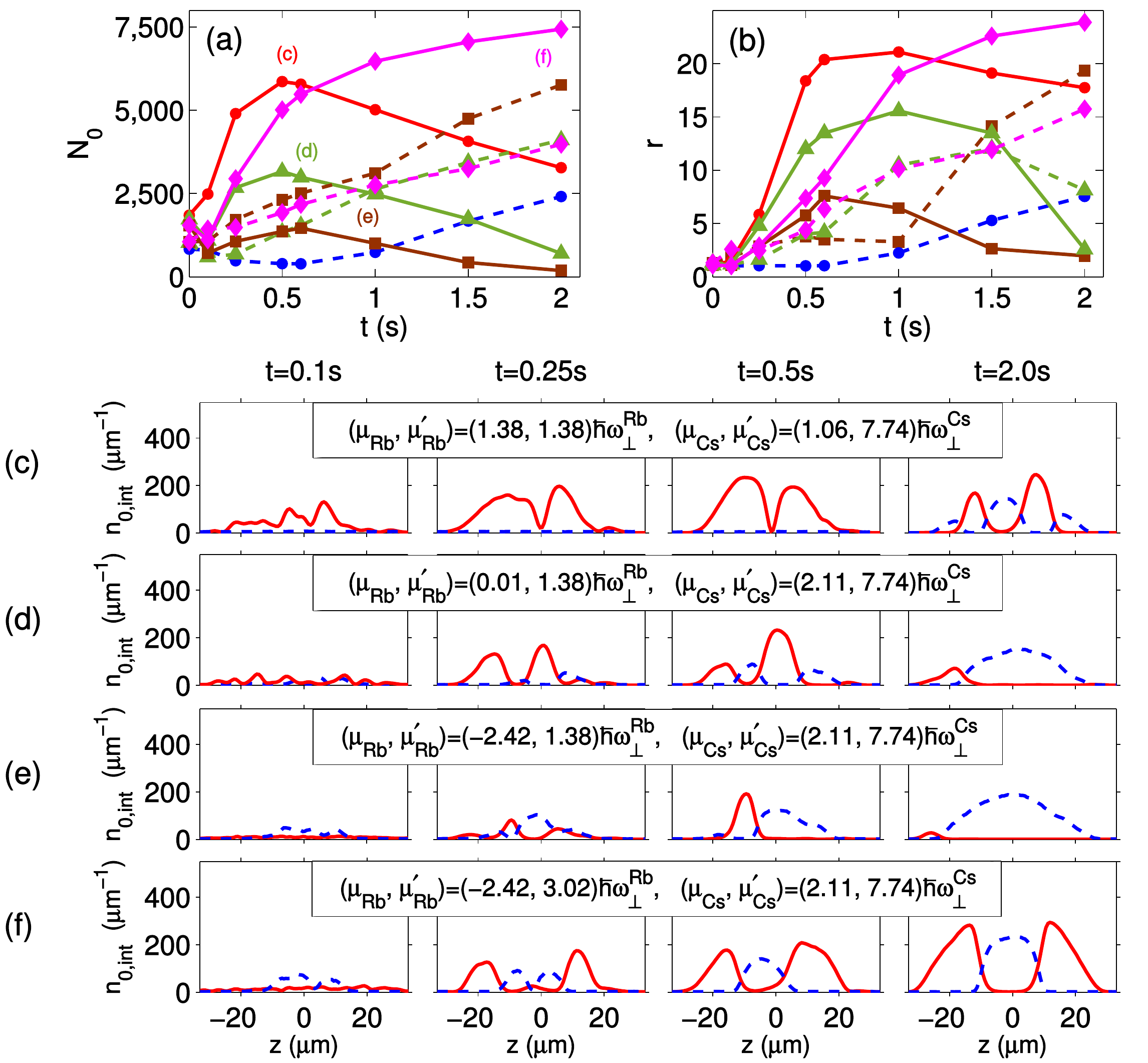}
\caption{(color online). Evolution of (a) condensate atom number, (b)
occupation ratio $r$, and (c)--(f) coupled 1D integrated condensate density
profiles for initial states with different amount of coherence in the Rb
initial state. Shown are cases with $\mu_{\rm Cs}>0$ and (c) $\mu_{\rm Rb}=1.38
\hbar \omega_\perp^\mathrm{Rb}$ (corresponding to ``reference''
Fig.~\ref{fig:2} data, plotted only for ease of comparison here), (d) $\mu_{\rm
Rb}=0.01 \hbar \omega_\perp^\mathrm{Rb}$, and (e)--(f) $\mu_{\rm Rb}=-2.42\hbar
\omega_\perp^\mathrm{Rb}$; (f) differs from (e) only in the larger post-quench
Rb chemical potential [$\mu_{\rm Rb}' = 3.02\hbar \omega_\perp^\mathrm{Rb}$ in
(f), as opposed to $1.38\hbar \omega_\perp^\mathrm{Rb}$ in (e)]. Note that for
(d)--(f) $\mu_{\rm Cs}$ is increased to $2.11\hbar \omega_\perp^\mathrm{Cs}.$
\label{fig:5}}
\end{figure}

{\em Quench Protocol and Modeling Details: }
We analyse a sudden temperature and chemical potential quench of a prolate
(aspect ratio $\approx 10$) two-component atomic cloud of approximately $1.4
\times 10^{6}$ $^{87}$Rb and $8 \times 10^{5}$ $^{133}$Cs atoms, equilibrated
in slightly displaced traps at $T_0=80$~ nK (close to the ideal gas critical
temperature $T_{\rm c}$).  Based on a characteristic example
(Fig.~\ref{fig:2}), in which neither component initially possesses a single
macroscopically-occupied mode, we analyze the role of fluctuations and quench
parameters [Figs.~\ref{fig:3}, \ref{fig:4}(a)], under the assumption that Rb
condenses fastest.  We then analyze a range of other initial states and
quenches, and critically discuss our main assumptions [Figs.~\ref{fig:4}(b),
\ref{fig:5}, and \ref{fig:2}(f,~inset)]. Within current computational
constraints, this provides a qualitative characterisation of the dynamical
phase diagram.

Dynamical two-component BEC simulations to date have been based on coupled
ordinary \cite{Busch2001a,Ohberg2001a,Kasamatsu2006a,Sasaki2011a} or
dissipative \cite{Ronen2008a,Achilleos2012a,pattinson_2014} Gross-Pitaevskii
equations (GPEs), classical field \cite{Berloff2005a}, truncated Wigner
\cite{Sabbatini2011a,Sabbatini2012a}, or ZNG (coupled GPE-Boltzmann)
\cite{edmonds_lee_15} methods.  The effects of thermal fluctuations during
condensate growth are best captured by 3D coupled stochastic projected
Gross-Pitaevskii equations \cite{blakie_bradley_08,bradley_blakie_14,De2014a}, 
\begin{equation}
      \total\psi_{k}=\mathcal{P}\left\{-\frac{i}{\hbar}L_{k}\psi _{k}+\frac{\gamma_{k}}{k_{\rm B}T}\left(\mu_{k}-L_{k}\right)\psi _{k}\right\}\total t+\total W_{k},
      \label{eq:SPGPE}
\end{equation}
describing the evolution of highly-occupied ``classical'' modes $\psi_k$, where
$k$ labels the species (here Rb or Cs). Here $L_{k}=-\hbar^{2}\nabla^{2}/2m_{k}
+ V_{k} + 4\pi \hbar^2 (a_{kk}\left|\psi_{k}\right|^{2}/m_k +
a_{kj}\left|\psi_{j}\right|^{2}/M_{kj})$, for scattering lengths $a_{kk}$
(intraspecies) and $a_{kj}$ (interspecies), atomic masses $m_k$, and reduced
mass $M_{kj}$.  The ``c-field'' region defined by projector $\mathcal{P}$,
consists of single-particle modes $\phi _{l}$ with energies below a
carefully-selected energy cutoff \footnote{In order to make our simulations
numerically tractable, we use a plane wave basis without strict dealiasing (a
procedure subtly different from Refs.~\cite{blakie_bradley_08,Blakie2008b}).
However, we have confirmed that our main results remain qualitatively similar
when the cutoff is changed by $\sim 10\%$.}, which is coupled to the
above-cutoff reservoir at temperature $T$ via growth (described by the
species-dependent rates $\gamma_{k}$) and noise (described by $\total W_{k}$
\footnote{Here $\total W_{k}$ is complex white noise defined by
\unexpanded{$\left\langle \total W_{k}^{\ast }\left( \mathbf{x},t\right) \total
W_{j}\left( \mathbf{x}^{\prime },t\right) \right\rangle =2\gamma _{k}\delta
_{kj}\delta _{C;k}\left(\mathbf{x-x}^{\prime }\right) \total t$}, where
\unexpanded{$\delta _{C;k}\left( \mathbf{x-x}^{\prime }\right)
=\sum_{l}^{E_{l}\leq E_{c;k}}\phi _{l}^{\ast }\left( \mathbf{x}^{\prime
}\right) \phi _{l}\left(\mathbf{x}\right)$} .}) processes. To simulate
two-component condensate formation, we first obtain an initial
state by numerically propagating Eq.~(\ref{eq:SPGPE}) to equilibrium for
initial chemical potentials $\mu_{k}$ and common temperature $T_{0} \gtrapprox
T_{\rm C}$, and then instantaneously quench these parameters to final values
$T<T_{\rm C}$ and $\mu_{k}'\geq\mu_{k}$. We extract the time-dependent
condensate mode using the Penrose--Onsager (PO) criterion \cite{Penrose1956a},
by seeking a macroscopically-occupied eigenmode of the one-body density matrix
in short-time averages over single trajectories \cite{blakie_bradley_08}.

{\em Typical Coupled Dynamics Simulation: }
A typical numerical run of the post-instantaneous-quench dynamics demonstrating
the formation of a composite defect for a given initial state (see figure
caption for parameters) is shown in Fig.~\ref{fig:2}, a result with which
subsequent results will be compared.  Snapshots (left panels) of the c-field
density (a), condensate mode density (b), and condensate phase (c) reveal three
characteristic dynamical stages: (i) a short post-quench ``condensation onset''
stage with strongly non-equilibrium dynamics (for $\lesssim 0.1\,$s) in which
multiple quench-induced defects proliferate; (ii) a rapid relaxation stage
(here $t\lesssim 0.5\,$s, featuring rapid Rb growth), dominated by defect
coarse-graining dynamics, leading to a metastable equilibrium state with a
number of long-lived defects [this number can be zero, one (as shown here), or
potentially more]; (iii) a slow evolution towards global equilibrium, during
which the other component (here Cs) may (but need not) also condense.  While
the details of the coupled dynamics are sensitive to global
system and quench parameters, our subsequent analysis confirms that the
occurrence of these three stages is generic.  Within our model, for a given set
of quench parameters, the long-term evolution tends to favor the dominance of
{\it either} Rb or Cs.

The right panels in Fig.~\ref{fig:2} show the dynamics of the PO condensate
atom number, $N_0$ (d), the occupation ratio, $r$, of the condensate mode to
the next-most-highly occupied mode (e), and the number of soliton defects (f).
While initially $r \sim 1$, it increases rapidly post-quench as a
macroscopically-occupied condensate mode emerges. Coherence growth, caused by
the suppression of phase fluctuations during condensation
\cite{hugbart_retter_07,esslinger_07a}, coincides with the rapid decrease in
the defect numbers due to merging and decay \cite{trento_prl}. Due to the
larger initial Rb atom number, Cs condensation follows that of Rb
\cite{Papp2008a}; moreover, as our simulations do not include evaporative
cooling, we find that Cs condensation is associated with a gradual decrease in
the Rb condensate.

Dependent on dimensionality \cite{delcampo_kibble_13}, the emerging defect
could also be a vortex or solitonic vortex \cite{Brand2002a}, whose infilling
would create a (solitonic) vortex--bright soliton \cite{Law2010a} rather than a
dark-bright soliton.  However, our geometry and parameters, which have
$\{\mu_\mathrm{Rb}, \mu_\mathrm{Rb}'\} < 4 \hbar \omega_\perp^\mathrm{Rb}$ and
$\{\mu_\mathrm{Cs}, \mu_\mathrm{Cs}'\} < 8 \hbar \omega_\perp^\mathrm{Cs}$,
favor 1D solitonic defects \cite{Brand2002a}. In Fig.~\ref{fig:2}, the initial
multiple defects in the fastest condensing component eventually lead to exactly
one {\em long-lived} defect, which acts as a tight mobile microtrap
\cite{Stamper-Kurn1998a}, facilitating rapid condensation of the other
component.  Growth of the bright (infilling) component in the microtrap (which
stabilizes the dark soliton against decay \cite{Achilleos2012a}) leads to a
spontaneously-generated dark-bright soliton; for this particular simulation,
this soliton forms close to the trap centre.  Over longer timescales, the
growing mean-field potential converts this dark-bright solitary wave into a
fixed domain wall \cite{Coen2001a,Stamper-Kurn1999a}.  A Cs condensate also
subsequently forms at the system edges, slightly compressing the Rb condensate
towards the trap center.  This interplay between condensate growth, defect
formation and dissipative evolution in the first component, and condensate
growth in the second component --- alongside increasing mean-field repulsion
between the co-forming immiscible condensates --- leads to interesting features
that should be observable in current experiments.  In the following we
consider: (i) the role of shot-to-shot fluctuations (Fig.~\ref{fig:3}); (ii)
dependence on growth rates and final quench parameters when starting from
quasi-condensate initial states (Fig.~\ref{fig:4}); and (iii) changes to the
evolution due to less coherent initial states (Fig.~\ref{fig:5}) and
finite-duration quenches [Fig.~\ref{fig:2}(f, inset)].  Importantly, we find
that the scenario of a defect in Rb leading to Cs infilling can be effectively
reversed for particular parameters and timescales [Fig.~\ref{fig:5}(f),
$t=0.25$s].

{\em Shot-to-Shot Variations: } We investigate sensitivity to fluctuations by
looking at different numerical runs (loosely corresponding to different
experimental runs) for the same fixed quench sequence
[Fig.~\ref{fig:3}(a)--(d)]. Such runs feature statistically-identical
condensate number evolution [see error bars in Fig.~\ref{fig:4}(a)].
Figure~\ref{fig:3}(a) reveals a three-stage evolution in the ``reference'' case
of Fig.~\ref{fig:2}: slow-moving solitonic defect, near-stationary dark-bright
solitary wave, and eventual static domain wall structure.
Figures~\ref{fig:3}(b)--(d) reveal alternative (and roughly equally-likely)
outcomes for different (same mean amplitude) post-quench dynamical noise
realizations, describing thermal fluctuations.  Figure~\ref{fig:3}(c) shows
slower Cs growth, in an initially shallower (hence more rapidly moving) dark
solitary-wave microtrap. Unlike Fig.~\ref{fig:3}(a), where dark solitary wave
decay and stabilisation occurs at the trap center, Figs.~\ref{fig:3}(c)--(d)
show this process occurring either to the left (c), or to the right (d),
illustrating the crucial role of the motion of the decaying dark solitary wave
\cite{Cockburn2010a}.  Figure~\ref{fig:3}(b) illustrates the potential for
co-existence of more than one (shallow) defect for an appreciable time [see
also Fig.~\ref{fig:4}(d) for $t \lesssim 0.5$s], which typically decay without
Cs growth in the solitonic microtraps. In this case, Cs only condenses around
Rb, minimizing mean-field repulsion.  Figure~\ref{fig:3}(e)--(h)
shows the condensate profile after 1 s (insets 2 s), consisting of either a
large Cs structure in the middle separating the Rb [inset to (e)], or a large
Rb structure enclosed by Cs [insets to (f), (g), (h)]. Further evolution of
case (a) (to $\gtrsim 3$s) also yields an Rb structure enclosed by Cs, after
gradual merging of the two Rb peaks.

We have confirmed that the qualitative findings above are unchanged when using
different (same temperature) thermalized initial states in the same
simulations, loosely corresponding to different experimental runs. In
particular, although different, randomly-generated, initial (pre-quench) states
yield different post-quench defect formation dynamics, we find that at least 3
out of 9 simulations with different initial and dynamical noise reveal
clear evidence of spontaneous dark-bright soliton formation, with 2 of the
emerging composite defects being relatively deep and central.  Despite the
differences in emergent density profile, condensate atom number evolution is
the same [within statistical variations, see error bars in
Fig.~\ref{fig:4}(a)], irrespective of the formation, stabilization or decay of
the dark-bright solitary wave.  The importance of dark-bright solitary waves in
the early stages of formation is further confirmed through dissipative GPE
simulations, showing that even perfectly-imprinted multiple dark-bright
solitons quickly coalesce into a single long-lived dark-bright soliton, with
rapid dynamical mean-field stabilisation. Importantly, we find that the
presence, or absence, of small asymmetries ($\sim$1~$\upmu$m) in the trap
minima only alter the details, but not the qualitative structure, of the
emerging density profiles, indicating that shot-to-shot and thermal
fluctuations can strongly suppress experimentally relevant trap imperfections
(whose role can dominate mean-field simulations \cite{Pattinson2013a}).

{\em Dependence on Global Quench Parameters: }
Distinct dynamical regimes occurring for different quench parameters are shown
in Fig.~\ref{fig:4}.  We focus on the dependence of $N_0$ on variations of the
chemical potentials $\mu_k'$ and growth rates $\gamma_k$ for fixed initial
quasi-condensate states. With $\gamma_{\rm Rb} = \gamma_{\rm Cs}$ fixed, we
find Rb dominates the evolution for a broad range of $\mu_{\rm Rb}'>\mu_{\rm
Rb}$ [squares/black lines in Fig.~\ref{fig:4}(a)]. Here,
spontaneously-generated defects tend to decay within Rb, with Cs not condensing
into a dark-bright solitary wave microtrap but instead at the trap edges
[Fig.~\ref{fig:4}(c)].  For quenches with large $\mu_{\rm Cs}'$
[Fig.~\ref{fig:4}(a), green triangles], despite the early atom number evolution
remaining practically unchanged, Cs condensation rapidly overtakes that of Rb,
which gradually disappears [Fig.~\ref{fig:4}(d)].  We have also investigated
the effect of the growth rate on the condensate evolution (standard estimates
\cite{blakie_bradley_08} suggest that $\gamma_{\rm Cs} \sim 10 \gamma_{\rm
Rb}$), and find that increasing $\gamma_{\rm Cs} / \gamma_{\rm Rb}$, for fixed
chemical potentials, favours the rapid growth of Cs (after a brief initial
transient of rapid Rb growth) [Fig.~\ref{fig:4}(b)].

{\em Role of Quasi-Condensation in Pre-Quench States: }
Our simulations so far exhibit a consistent predominance of rapid Rb growth
{\em initially}, set by the pre-quench conditions, with long-term evolution
dictated by the quench parameters. Although our initial states featured no
single macroscopically-occupied state in either component, both initial states
exhibited quasi-condensation, with multiple states having occupations greater
than half that of the most-occupied state [$\approx 10$ states for Rb (largest
occupation $\approx 2200$), and $\approx 5$ states for Cs (largest occupation
$\approx 1100$)].  To demonstrate the generality of our findings,
Fig.~\ref{fig:5} shows the effects of decreasing the initial Rb
quasi-condensation, i.e.\ changing $\mu_{\rm Rb}$  from $>0$
[Fig.~\ref{fig:5}(c)] to $\approx 0$ [Fig.~\ref{fig:5}(d)] and $<0$
[Fig.~\ref{fig:5}(e)], while allowing for an initial, more pronounced, Cs
quasi-condensate (through the increased $\mu_{\rm Cs} = 2.1 \hbar
\omega_\perp^{\rm Cs}$).  This leads to a large Cs condensate forming before
significant Rb condensate growth.  Under these conditions we find that
enhancing the final Rb chemical potential, $\mu_{\rm Rb}'$
[Fig.~\ref{fig:5}(f)], can lead either to approximately equal Rb and Cs
condensate number, or to the generation of a quasi-stable Cs defect enabling
the short-time formation and entrapment of an Rb condensate (at $t=0.25$s).

{\em Effects of Finite Quench Duration: }
As a further check on the generality of our results, we also consider
finite-duration quenches in which  ($T_k$, $\mu_k$) are ramped linearly from
the same initial to final values as in Fig.~\ref{fig:2} over time $t_R$.
Provided the ramp duration does not exceed the characteristic timescale for the
emergence of a composite defect (i.e., $t_{R} \lesssim 500$ms) our qualitative
findings remain practically unaffected; we still observe, in some stochastic
trajectories, spontaneous dark-bright solitary wave defects which can influence the
later density profiles.  For these finite-duration quenches we observe scaling
of the immediate post-quench defect number with $t_R$ shown in
Fig.~\ref{fig:2}(f). There is some qualitative resemblence to scalings observed
in the short- and long-term evolution of spinor gases quenched from miscible to
immiscible in homogeneous or harmonically-trapped systems~\cite{Sabbatini2011a,
Sabbatini2012a, Swislocki2013a, Swislocki2015a, Hofmann2014a}; however, the
lack of a clear KZ scaling in our data is expected, since the presence of
quasi-condensation in our pre-quench states in particular does not conform to
the adiabatic--impulse--adiabatic scenario of the KZ mechanism.

{\em Experimental Relevance:}
Recent experiments have explored the emergent immiscible density profiles
following the gradual sympathetic cooling of $^{87}$Rb-$^{85}$Rb
\cite{Papp2008a}, or $^{87}$Rb-$^{133}$Cs \cite{McCarron2011a} mixtures.
Although our simulations feature sudden quenches, we can still draw some
interesting, albeit indirect, analogies to these experiments: (i) Both
experiments sometimes found (depending on parameters) a condensate in only one
component, in qualitative agreement with the long-term evolution of our
stochastic simulations.  (ii) Our simulations reveal long-lived metastable
phase-separated profiles resembling those seen in experiment
\cite{McCarron2011a}. While the energy of the immiscible system is generally
lowered by reducing the number of domains, larger aspect ratios add an energy
barrier to the process of reduction; the experiment of Ref.~\cite{Papp2008a}
(with larger aspect ratio) reported metastable states with lifetimes exceeding
1s, consistent with our simulation timescales.  (iii) The relative timescales
for condensation in the two components influences the form of the emergent
profiles, as the component condensing second (typically with fewer initial
atoms, and thus lower critical temperature) can only do so where the density of
the other component is low \cite{Papp2008a,McCarron2011a}; this explains why,
over the probed timescale, simulations in the scenario of Fig.~\ref{fig:3} most
commonly exhibit a central Rb condensate surrounded by the more massive Cs (a
feature not easily reproduced at equilibrium \cite{Pattinson2013a}).  The
long-term evolution will, in principle, erase fluctuations from the formation
dynamics (although in practice this timescale may be too long to be
experimentally relevant). Indeed, longer-term evolution in the scenario of
Fig.~\ref{fig:2} ($\ge 3$s) exhibits coalescence of the two Rb structures into
a single structure surrounded by Cs. This coalescence often occurs shortly
after the Rb condensate number decreases below that of Cs, providing a
plausible indirect explanation for the observation of such a regime in
Ref.~\cite{McCarron2011a}.

{\it Conclusions: }
We have qualitatively analysed the formation dynamics of immiscible
two-component condensates following a sudden temperature quench, elucidating
the importance of composite defect formation and dynamics.  While our model is
computationally challenging (even the coupled dissipative model, without noise,
has 8 independent parameters), analysis of over 100 3D stochastic simulations
has enabled us to broadly classify the dynamics into 3 evolution stages: (i)
condensation of the fastest-growing component, with dynamics mainly determined
by the pre-quench state and final temperature, in which multiple defects are
spontaneously formed; (ii) fluctuation-determined coarse-graining dynamics,
during which defects in the fastest-growing component coalesce and gradually
decay, potentially yielding a small number of long-lived defects; and (iii)
relaxation, with long-term dynamics set by the quench parameters, during which
the second component condenses in regions of low mean-field density.
Importantly, a statistically non-negligible fraction ($\gtrapprox 30 \%$) of
simulations exhibit a composite defect, whose spontaneous formation and
subsequent evolution leads to stark shot-to-shot variations in the observable
phase-separated density profiles over a broad range of experimentally relevant
timescales.  By varying the initial state and quench parameters, we have
confirmed that both species may exhibit a defect-induced microtrap following an
instantaneous quench, demonstrating the generic nature of such features.

While the prevailing composite defect in our inhomogeneous system is a
dark-bright solitary wave, higher dimensional analogues are also possible, and
the related ``core condensation'' dynamics in the context of condensed matter
and high energy physics remain only partly understood even in uniform systems
\cite{kibble_njp,antunes_gandra_06}.  Controlled experiments with cold atoms
could help shed further light on this important problem by, e.g., statistically
analysing the evolution of density profiles following quenched evaporative
cooling of binary gases initially near the critical temperature.

We acknowledge funding from the UK EPSRC (Grant Nos. EP/K03250X/1,
EP/K030558/1). IKL and SCG were supported by the National Science Council,
Taiwan (Grant No. 100-2112-M-018-001-MY3). TPB was supported by the John
Templeton Foundation via the Durham Emergence Project
(\href{http://www.dur.ac.uk/emergence}{http://www.dur.ac.uk/emergence}).
%

\end{document}